\documentclass[prl,twocolumn,flafter]{revtex4}

\usepackage{graphicx}
\usepackage{flafter}

\usepackage{ccaption}
  \captionnamefont{\bfseries}
  \captiontitlefont{\small\rmfamily}
  \captionstyle{\raggedright}
  \captiondelim{ }
  \normalcaption

\begin{document}

\bibliographystyle{apsrev}


\title{Electric-field control of a hydrogenic donor's spin in a semiconductor}

\author{A. De}
\author{Craig E. Pryor}
\author{Michael E. Flatt\'e}

\affiliation{Department of Physics and Astronomy and Optical Science
and Technology Center, \\ University of Iowa, Iowa City, Iowa 52242}

\date{\today}

\begin{abstract}
An AC electric field applied to a donor-bound electron in a semiconductor modulates the orbital character of its wave function, which affects the electron's spin dynamics via the spin-orbit interaction. Numerical calculations of the spin dynamics of a hydrogenic donor (Si) embedded in GaAs, using a real-space multi-band ${\bf k}\cdot {\bf p}$ formalism, show the high symmetry of the hydrogenic donor state results in strongly nonlinear dependences  of the electronic $g$ tensor on applied fields. A nontrivial consequence is that the most rapid Rabi oscillations occur for electric fields modulated at a {\it subharmonic} of the Larmor frequency.
\end{abstract}

\pacs{ 76.30.Da , 71.70.Ej , 76.30.-v , 71.55.-i }

\maketitle\vfill\eject


Electronic ground states characterized by non-zero spin are attractive candidates for encoding quantum information in a solid state system, and the use of electric fields is an attractive method to address individual spins\cite{Awschalom2002,Awschalom2007}.
When the ground state has a nonzero {\it integer} spin it is possible to perform all needed spin operations using electric fields alone\cite{Levy2002,Tang2006},
whereas for manipulation of spin-$1/2$ electronic ground states at least a static applied magnetic field is required.
Advances in focused ion beam single-ion implantation\cite{Shinada2005}, as well as the use of  a scanning tunneling microscope to implant a single ion with atom-scale precision\cite{Schofield2003,Kitchen2006},
suggest that spin clusters and spin-based circuits consisting of large numbers of precisely positioned spins could be designed with near-atomic resolution.
Proposals to control individual spin-$1/2$ states in such an environment with local electric fields include changing the magnitude of the Land\'e $g$ tensor to bring spins into resonance with an extended AC magnetic field\cite{Loss1998,Kane1998,Jiang2001,Nakaoka2007}, moving spins in a fringe-field\cite{Tokura2006}
or a hyperfine\cite{Petta2005} gradient, modulating zero-field spin splittings\cite{Rashba2003,Nowack2007}, and $g$-tensor modulation resonance ($g$-TMR)\cite{Kato2003}. $g$-TMR uses the electric-field dependence of the Land\'e $g$ tensor {\it anisotropy} to manipulate the spin, and so does not require  microwave magnetic fields or nanoscale magnetic materials or nuclear polarization gradients. Although $g$-TMR works by changing the orbital character of the wave function with an electric field, and thereby indirectly influencing the spin through the spin-orbit interaction, it does not require zero-field spin splittings (so $g$-TMR could be performed in a silicon or diamond host).   Predictions for quantum dots indicate control of the $g$ tensor anisotropy can produce rapid Rabi oscillations and full Bloch-sphere control with a single vertical electric field\cite{Pingenot2008}.

\begin{figure}
\includegraphics[width=0.9\columnwidth]{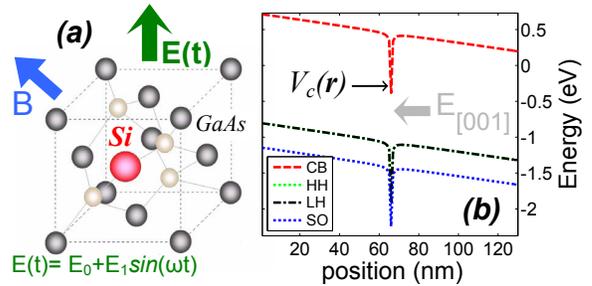}
\caption{ {\bf (a)} Proposed geometry for {\it g}-tensor modulation.
{\bf (b)} The impurity potential is  a Coulomb potential plus a
central cell correction. The effect of applying an electric field on
the zone center band energies is shown. } \label{fig1}
\end{figure}

The promising approach of $g$-TMR has yet to be explored for
electrons bound to dopants.  Shallow donors might seem a poor
candidate for modulation of $g$ tensor anisotropy, as they have
cubically-symmetric $g$ tensors in the absence of an electric field.
Quantum dots, by contrast, have highly asymmetric $g$ tensors that
are very sensitive to applied electric fields\cite{Pingenot2008}.
However, we find the $g$ tensors of electronic spins bound to donors
depend nonlinearly on applied electric and magnetic fields, and thus
substantial $g$ tensor anisotropy and rapid spin manipulation can be
achieved for a hydrogenic donor state. As the dominant
electric-field dependence is nonlinear, the most rapid Rabi
oscillations are found at unexpected frequencies --- subharmonics of
the Larmor frequency rather than the fundamental --- permitting
rapid spin manipulation using AC electric fields with frequencies
far below the Larmor frequency. Furthermore, the $g$ tensors of
quantum dots are very sensitive to dot shape and
composition\cite{Pryor2006b} and thus each quantum dot will have
different resonance frequencies for $g$-TMR. Donor wave functions
and $g$ tensors will, however, each be reliably the same.

These hydrogenic states have other attractive features for spin
clusters or spin devices; they possess the biggest radii of any
ionic bound states in the solid, with Bohr radii of the order of
10~nm in GaAs. Thus the spin-spin coupling between states would be
easier to control than for deep levels whose interaction strength
changes substantially on the atomic scale\cite{Kitchen2006}. Our
treatment focuses on the substitutional silicon donor in gallium
arsenide, ${\rm Si_{Ga}}$, as it is one of the best understood
semiconductor point defects and is well described by the hydrogenic
model. We expect that similar results are possible for a shallow
donor in silicon, although the details may be complicated by  the
presence of multiple valleys in the conduction band.

The geometry of $g$-TMR for a single electron spin bound to a donor
is shown in Fig. \ref{fig1}(a). A
static magnetic field along with a gated time varying electric field
is applied to the crystal containing the $\rm Si_{Ga}$ donor. We considered all orientations of the field and found that the
most rapid Rabi oscillations occur when the magnetic field is
applied at an angle $\theta=45^o$ to the electric field, which is the configuration shown in Fig.~\ref{fig1}(a).

Although many properties of shallow impurities (such as the energy spectrum) can be treated to an excellent approximation
by two-band effective mass theory\cite{Kohn1955},  $g$-tensor
calculations require a multi-band treatment as the coupling among
multiple bands needs to be considered, and the spin-orbit
interaction
must be treated accurately\cite{Roth1959}.
Moreover, the electric field breaks the
spherical symmetry of
the impurity site.
These complexities are best handled numerically.

Our calculations of $g$-TMR for $\rm Si_{Ga}$ donors were carried
out using 8-band  $\bf k \cdot p$ theory\cite{Bahder1990} in the envelope
approximation
using finite differences on a real
space grid \cite{Pryor1991, Pryor1998, Pryor1997, Pryor2006b}.
Material parameters were taken from
Ref.~\onlinecite{Vurgaftman2001} assuming $T=0$. The potential of
the hydrogenic impurity,
\begin{equation}
V_c( {\bf r})= \frac{e^2}{4\pi\epsilon{r}}+C\delta({\bf r}-{\bf
r_0}),
\end{equation}
is the sum of a screened Coulomb potential and a delta
function potential corresponding to the
central-cell-correction (CCC). The CCC arises due to the differing
chemical nature of various impurities.  For our
calculations $V_c({\bf r})$ is non-zero only on a single grid site, as
shown in Fig. \ref{fig1}(b) along with the potential due to the applied electric field.
The CCC is found by adjusting $C$ until the calculated binding
energy for the $1s$ donor state matches experiment. The Land\'e
$g$ tensor for the impurity ground state was then obtained from the calculated
Zeeman splitting of the $1s$ level in a uniform magnetic field.



\begin{figure}
\includegraphics[width=1\columnwidth]{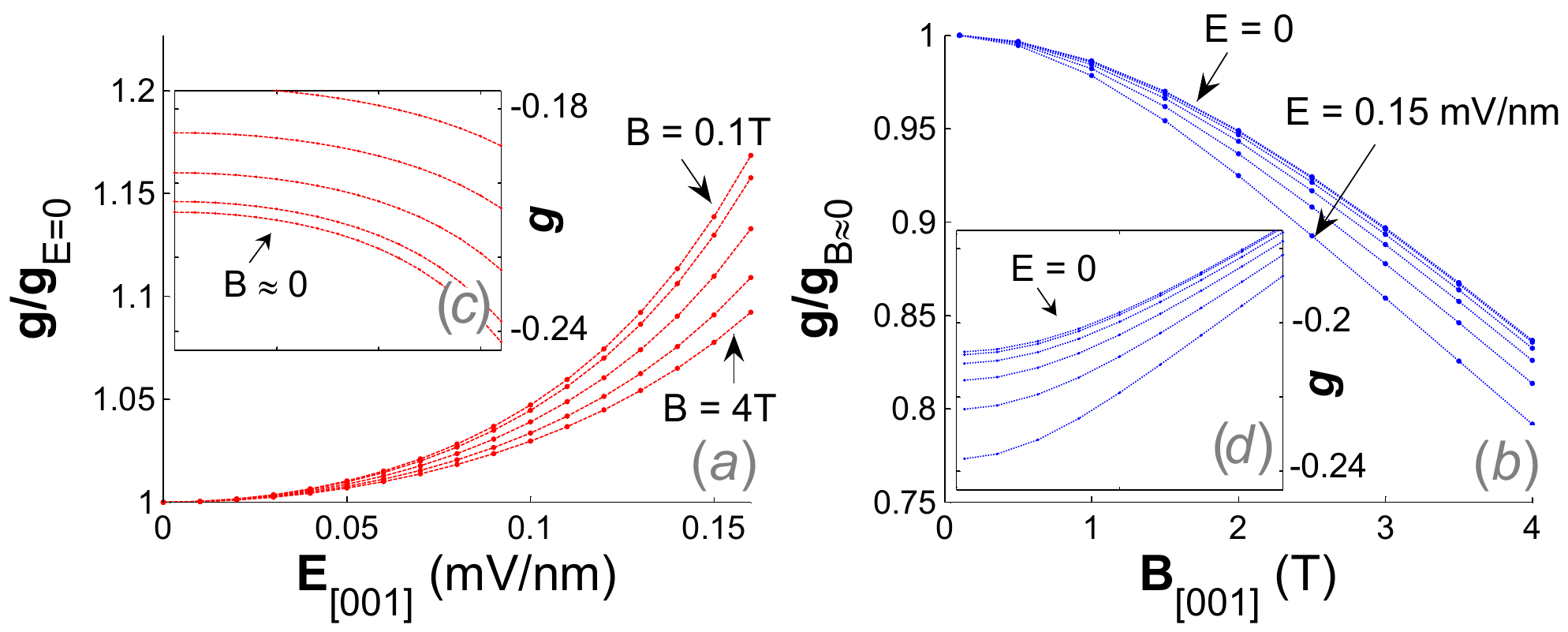}
\caption{ Normalized donor $g_{[001]}$ values as a function of (a)
${\bf E}_{[001]}$ and (b) ${\bf B}_{[001]}$. Insets show
unnormalized donor $g_{[001]}$ values. The full range of the
$x$-axis of the insets is
the same as that of their respective (normalized) plots. 
} \label{fig2}
\end{figure}

Fig.~\ref{fig2} shows $g_{[001]}$, the tensor component for the $1s$ impurity state  as a function of collinear magnetic and
electric fields.
 Increasing the electric field increases
the relative change in $g$, whereas increasing the magnetic field
decreases the relative change in $g$. Appreciable changes in $g$ are
seen even at modest magnetic fields, which is encouraging for
manipulating the donor atom's spin. The impurity $g_{[001]}$'s
depend nonlinearly on the magnetic field, as shown in
Fig.~\ref{fig2}(b). This behavior is unlike that seen in small QDs such as treated in Ref.~\onlinecite{Pingenot2008}, for which
the $g$ tensor is nearly independent of the applied magnetic
field.

\begin{figure}
\includegraphics[width=1\columnwidth]{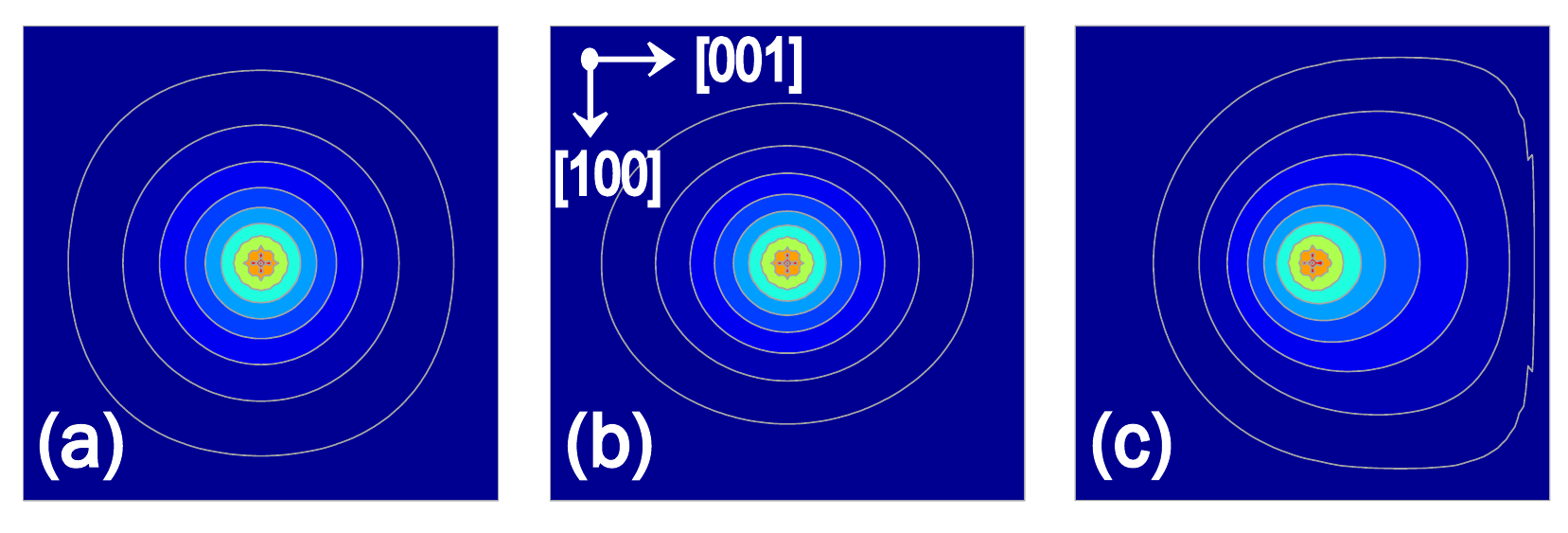}
\caption{ Calculated wavefunctions for {\bf (a)} ${E_{[001]}}=
B_{[001]}= 0$, {\bf (b)} $E_{[001]}$= 0, $B_{[001]}$= 4 T {\bf (c)}
$E_{[001]}= 0.15$~mV/nm, $B_{[001]}$= 0. Contours outline selected
amplitudes as a guide to the eye.}\label{fig3}
\end{figure}

The competing effects of $B$ and $E$ on $g_{[001]}$ can be
understood by examining the donor electron's wavefunction, shown in
Fig.~\ref{fig3}. As the magnetic field is increased in the $[001]$
direction [from Fig.~\ref{fig3}(a) to (b)], the cyclotron radius
decreases, contracting the extent of the wave function in the
direction transverse to ${\bf B}$. Similarly, the opposite effect is
evident when the electric field is increased [Fig.~\ref{fig3}(c)],
which allows the impurity wavefunction to spread into a region
with lower overall potential. This decreases the
confinement for the donor electron and thereby increases
$|g|$\cite{Pryor2006b}. This effect is more prominent for a
smaller magnetic field.

\begin{figure}
\includegraphics[width=1\columnwidth]{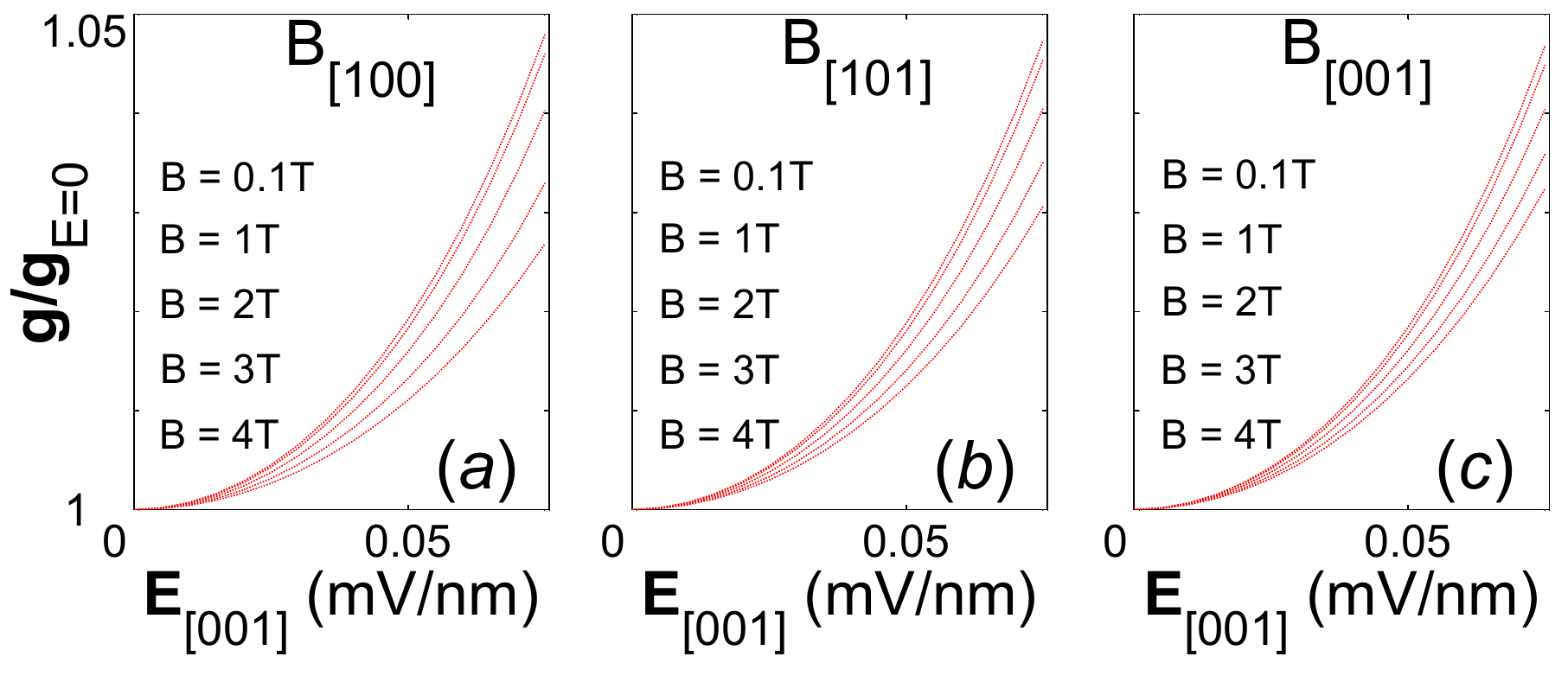}
\caption{ $g$ as a function of $E_{[001]}$ and $B$ applied in
various directions {\bf (a)} [100] {\bf (b)} [101] {\bf (c)} [001].
$g$ is also calculated for various magnetic field strengths. }
\label{fig4}
\end{figure}

The $g$ tensor components were calculated for various directions of
${\bf B}$ with ${\bf E}$ applied along [001], as shown in
Fig.~\ref{fig4}. Note that $\partial{g}/\partial{E}$ decreases with
increasing $B$. The variation in $\partial{g}/\partial{E}$ as a
function of $B$ is greater when ${\bf E}\perp {\bf B}$. However at
an intermediate $B$($\approx 2T$), $\partial{g}/\partial{E}$ is
identical in all directions. These
results imply that an
electric field induces a $g$ tensor anisotropy oriented relative to
${\bf E}$, which makes it possible to modulate the $g$ tensor using
an alternating electric field in addition to the static electric and
magnetic fields.

We next solve for the donor atom's spin dynamics by
explicitly integrating the time-dependent Schr\"odinger equation.
The nonlinear nature of $g$
complicates a {\it{quantitative}} treatment
within the rotating-wave approximation. The directional dependence
of $g$ (Fig.~\ref{fig4}) can be used to obtain an analytical
form of the $g$ tensor by fitting each tensor component to the
expression  $\displaystyle\sum_{n=0}^2{a_n({\bf
B})E^{2n}}$. A time-dependent $g$ tensor can then be constructed
for the time-varying electric field $E(t) = E_{dc} +
E_{ac}\sin(\omega{t})$. The maximum amplitude of $E(t)$ is always
held constant at $0.2$~mV/nm, so as not to exceed the
breakdown field of the GaAs host. The spin dynamics of the donor atom can
then be calculated using the effective time-dependent Hamiltonian,
\begin{eqnarray}
H(t) = \frac{\mu_B}{2}{\bf \sigma}\cdot{\bf\tilde g}(t)\cdot{\bf B}
\label{Ho}
\end{eqnarray}
where $\mu_B$ is the Bohr magneton.

\begin{figure}
\includegraphics[width=1\columnwidth]{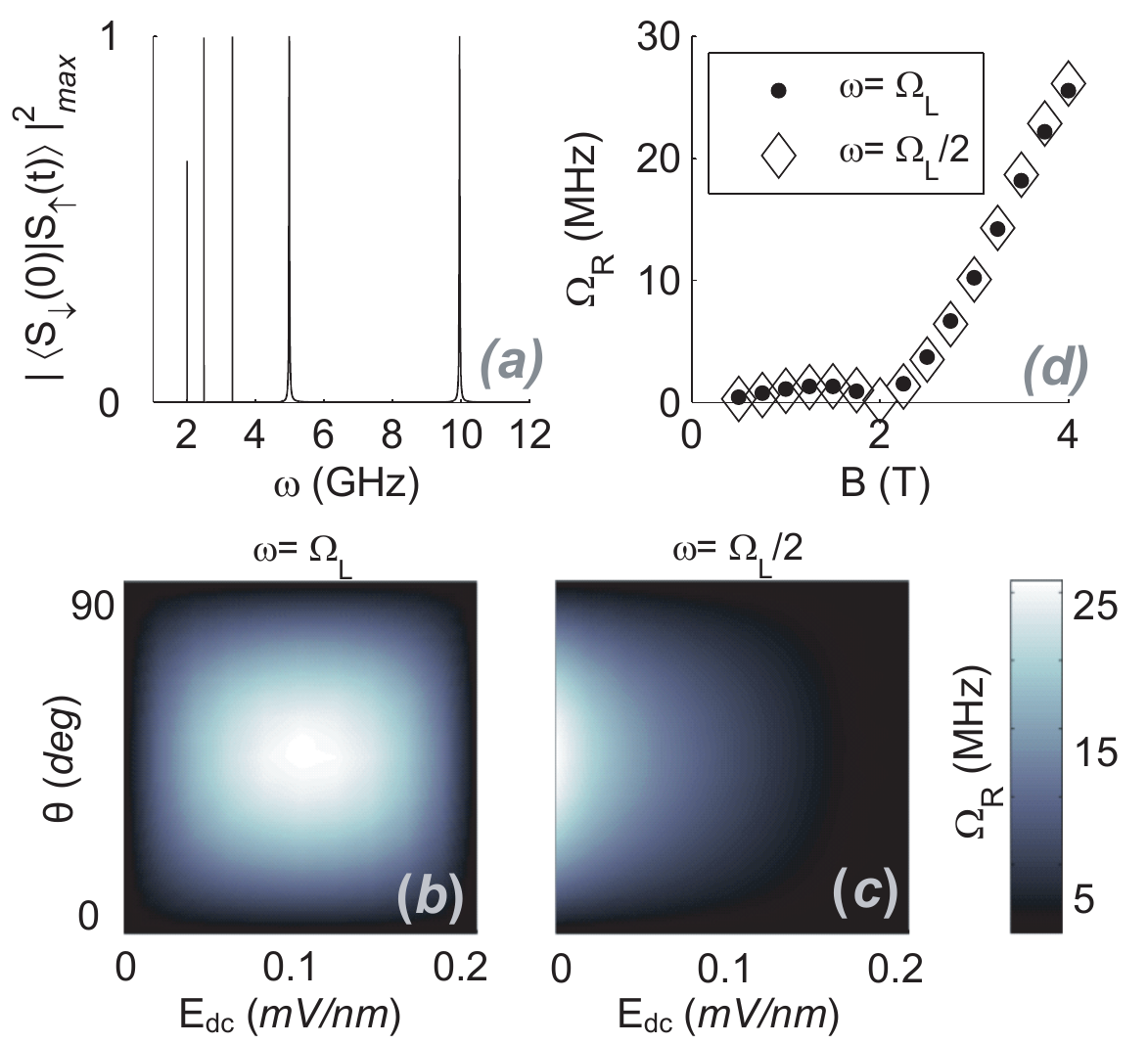}
\caption{ Spin dynamics of the donor atom as a function of various
parameters. $E_{ac}+E_{dc}=0.2$~mV/nm and is $[001]$ oriented.
$\theta$ is the angle between ${\bf B}$ and ${\bf E}$. {\bf (a)}
Peak spin-flip transition amplitudes as a function of $E$'s driving
frequency, for $E_{ac}/E_{dc}=9$ and $\theta=45^o$. Resonant
transitions appear at sub-harmonics of the Larmor frequency
$\Omega_L$. {\bf (bc)} Rabi frequency $\Omega_{R}$ as a function of
$E_{dc}$ and $\theta$ for: {\bf (b)} $E$ driven at $\Omega_L$,
$\Omega_{R}$ is maximum at $\theta=45^o$ and $E_{dc}=0.1$~mV/nm.
{\bf (c)} $E$ driven at $\Omega_L/2$, $\Omega_{R}$ is maximum at
$\theta=45^o$ and $E_{dc}=0$. {\bf (d)} $\Omega_{R}$ as a function
of $B$ for optimal $\theta$ and $E_{dc}$ of {\bf(b)} and {\bf(c)}.
Note that above $B=2T$,  $\Omega_{R}$ increases monotonically. }
\label{fig5}
\end{figure}

As the Hamiltonian is explicitly time dependent, the state of a
spinor, $S_j$ (where $j = \downarrow, \uparrow $) at time $t$ can be
obtained by evolving $S_j(t=0)$ forward in time in $n$ steps of $\Delta{t} =t/n  \ll 1/{2\omega}$ as follows,
\begin{eqnarray}
|S_j(t)\rangle=
{\bf\displaystyle\hat{T}}\displaystyle\prod_{\nu=0}^{n}
\exp\left(\frac{iH(t_\nu)\Delta{t}}{\hbar}\right) |S_j(0)\rangle
\label{T1}
\end{eqnarray}
where $\bf\hat{T}$ is the time-ordering operator.  For sufficiently small $\Delta{t}$ this is equivalent to
\begin{eqnarray}
|S_j(t)\rangle= {\bf\displaystyle{\hat
T}}\exp{\displaystyle\int_{0}^{t}}\left(\frac{iH(t')d{t'}}{\hbar}\right)
|S_j(0)\rangle, \label{T2}\end{eqnarray}
 The time
dependent probability of making a spin-flip transition is
$|\langle{S_{\uparrow}(0)}|{S_{\downarrow}(t)}\rangle|^2$.
 Rabi oscillations are obtained when spin flip
transitions are made resonantly
($i.e.~|\langle{S_{\uparrow}(0)}|{S_{\downarrow}(t)}\rangle|^2_{max}=1$).
Resonant spin flip transitions are usually made when $E(t)$ is
driven at the Larmor frequency $\Omega_L$. However in case of the
hydrogenic impurity system considered here, the donor electrons spin
can be resonantly flipped at any sub-harmonic of the Larmor
frequency: $\Omega_L/N$, where $N$ is an integer. This is
illustrated in Fig.~\ref{fig5}(a), where the peak spin-flip
transition probabilities are shown as a function of the driving
$E$-field frequency $\omega$. Multiple resonance lines are apparent,
located at $\Omega_L$ and its sub-harmonics. This unusual behavior
arises from the highly nonlinear dependence of ${g}$ on the applied
electric field (Fig.~\ref{fig4}). For sub-harmonics higher than
$N$=2, the Rabi frequencies $\Omega_R$ are lower than those at $N<2$
and hence are not considered further for spin manipulation. The
largest $\Omega_R$ can be achieved by driving $E$ at the second
sub-harmonic ($N=2$) of $\Omega_L$. Due to the smaller DC component
of the electric field the Rabi oscillations are less rapid at
$\Omega_L$, than at its second sub-harmonic. The resonance lines in
Fig.~\ref{fig5}(a)  at $\omega = \Omega_L / N$, have a full width at
half maximum of $\Delta \omega = 2 \Omega_R / N$.

The Rabi frequencies are calculated next as a function of $E_{dc}$
and $\theta$ and are shown in Fig.~\ref{fig5}(b) with the electric
field driven at $\Omega_L$. For all $\theta$, and $\omega=\Omega_L$,
$\Omega_R$ is largest when the AC and DC components of the electric
field are equal. If the electric field is driven at $\Omega_L/2$,
however, as shown in Fig.~\ref{fig5}(c), then $\Omega_R$ is largest
if $E_{dc}=0$. In both Figs.~\ref{fig5}(b) and (c), the optimal
angle of the magnetic field  to the electric field is $\theta=45^o$.
Although the maximum $\Omega_R$ in Figs.~\ref{fig5}(b) and (c) are
identical, driving $E$ at $\Omega_L/2$ offers two key advantages.
When the peak value of $E$ is close to the breakdown of the host
crystal, a pure AC field with an adjustable duty-cycle is much less
likely to ionize the donor electron, as the carriers can recover
during a thermal relaxation time. This allows for higher driving
fields, which result in higher $\Omega_R$. It also may be
experimentally more feasible to resonantly flip the spin at the
lower frequency of the subharmonic $\Omega_L/2$ than the fundamental
$\Omega_L$.

Fig.~\ref{fig5}(d) shows $\Omega_R$ as a function of $B$ for
$\theta=45^o$ and $E$ driven at $\Omega_L$ or $\Omega_L/2$. For
magnetic fields greater than $2T$, $\Omega_R$ increases
monotonically, whereas below $2T$ $\Omega_R$ exhibits a
non-monotonic behavior. This feature can be explained by
Taylor-expanding the time-dependent Hamiltonian to first order in
the rotating wave approximation,
\begin{eqnarray}
H(E) \approx \frac{\mu_B\bf\sigma}{2}\cdot\left({\bf\tilde g}+
\frac{E_{ac}}{2}\frac{\partial\bf\tilde g}{\partial
E}|_{E=E_{dc}}\right)\cdot\bf{B}.
\end{eqnarray}
Here the Larmor frequency is given by the time independent
static precession vector, ${\bf\Omega_0} ={\mu_B}{}{\bf\tilde
g}\cdot{\bf B}/\hbar$ and the electron's spin dynamics in the
rotating frame is described by the time-dependent spin precession
vector, ${{\bf\Omega}_1(t)} ={\mu_B}{}E_{ac}({\partial\bf\tilde
g}/{\partial E})\cdot{\bf B}/2\hbar$. ${\bf\Omega}_1$ can be
resolved into components that are parallel (${\bf\Omega_{||}}$) and
perpendicular (${\bf\Omega}_{\perp}$) to ${\bf\Omega}_0$. In the
rotating frame, $|{\bf\Omega}_{\perp}|$ is equivalent to $\Omega_R$
(in the lab frame), as driving $E$ at $|{\bf\Omega}_0|$ leads to
spin precession about ${\bf\Omega}_{\perp}$ or Rabi oscillations. As
the tensor components $\partial{{g}}/\partial{{E}}$
decrease with increasing $B$ (see Fig.~\ref{fig4}), the magnitudes
of $B$ and $\partial{{g}}/\partial{{E}}$ have opposing effects on
$\Omega_1$ (and hence $\Omega_{\perp}$). For $B<1$T the contribution
from $\partial{g}/\partial{E}$ dominates over $B$ and hence the Rabi
frequencies increase. For $1$T$<B<2$T the competing contributions of
$B$ and $\partial{{g}}/\partial{{E}}$ make the $g$ tensor
increasingly isotropic and the Rabi frequencies smaller.
At $B\approx2$T the $g$ tensor becomes isotropic and the Rabi frequency vanishes. Above $B \approx
2T$, the effects of a much larger magnetic field dominate and the
spin flip times decrease monotonically. Two key inferences, consistent with other work on $g$-TMR, can be
drawn from this behavior. For spintronic applications the highest
magnetic field possible is desirable in order to generate the largest possible Rabi
frequencies. Secondly,  the amount of $g$-tensor anisotropy
induced is crucial to achieving shorter spin-flip times, not the degree of change in $g$ as a function of ${E}$.


We have proposed a scheme for achieving electric-field driven
$g$-tensor modulation resonance for a single shallow donor impurity.
Electric and magnetic field dependent \hbox{$g$ tensors} were calculated
for the $\rm Si_{Ga}$ donor  using 8-band ${\bf k}\cdot {\bf p}$ theory on a real
space grid. Varying ${\bf E}$ and ${\bf  B}$ affects the confinement
for the donor electron, which in turn alters its $g$ tensor. In
addition to the nonlinear $E$ dependence, the $g$ tensors are also
highly nonlinear as a function of $B$. This is unlike the case for a
QD, where $g$ is essentially independent of $B$.  A consequence of
the nonlinear dependence of $g$ on $E$ is that spin-flip transitions
can be made resonantly at any sub-harmonic of the Larmor frequency.
Spin flip times were calculated exactly, using time evolution
operators, and optimized for various parameters of interest. If $E$
is driven at the second sub-harmonic of the Larmor frequency, then
high frequency Rabi oscillations can be obtained without any DC
component of $E$. This could be particularly useful in obtaining the largest Rabi frequencies for a given breakdown-field limit for the semiconductor host.

C.E.P. would like to acknowledge a NSF NIRT.
M.E.F. would like to acknowledge an ONR MURI.

\end{document}